# Cross-Scale Sensitivity Analysis of a Non-Small Cell Lung Cancer Model: Linking Molecular Signaling Properties to Cellular Behavior

**Zhihui Wang** [1], **Christina M. Birch** [2], **and Thomas S. Deisboeck** [1,§]

[1] Harvard-MIT (HST) Athinoula A. Martinos Center for Biomedical Imaging, Massachusetts General Hospital, Charlestown, MA 02129, USA; [2] Department of Biochemistry and Molecular Biophysics, University of Arizona, Tucson, AZ 85721, USA.

**Running Title:** Sensitivity Analysis of a NSCLC Model
**Abbreviations:** *EGF* = epidermal growth factor; *EGFR* = EGF receptor; *ERK* = extracellular signal-regulated kinase; *MAPK* = mitogen activated protein kinase; *MEK* = mitogen activated protein kinase kinase; *PLCγ* = phopholipase Cγ; *PKC* = protein kinase C.

[§]**Corresponding Author:**

Thomas S. Deisboeck, M.D.
Complex Biosystems Modeling Laboratory
Harvard-MIT (HST) Athinoula A. Martinos Center for Biomedical Imaging
Massachusetts General Hospital-East, 2301
Bldg. 149, 13th Street
Charlestown, MA 02129
Tel: 617-724-1845
Fax: 617-726-7422
Email: deisboec@helix.mgh.harvard.edu

## ABSTRACT

Sensitivity analysis is an effective tool for systematically identifying specific perturbations in parameters that have significant effects on the behavior of a given biosystem, at the scale investigated. In this work, using a two-dimensional, *multiscale* non-small cell lung cancer (NSCLC) model, we examine the effects of perturbations in system parameters which span both molecular and cellular levels, i.e. *across* scales of interest. This is achieved by first linking molecular and cellular activities and then assessing the influence of parameters at the molecular level on the tumor's spatio-temporal expansion rate, which serves as the output behavior at the cellular level. Overall, the algorithm operated reliably over relatively large variations of most parameters, hence confirming the *robustness* of the model. However, three pathway components (proteins PKC, MEK, and ERK) and eleven reaction steps were determined to be of *critical* importance by employing a sensitivity coefficient as an evaluation index. Each of these sensitive parameters exhibited a similar changing pattern in that a relatively larger increase or decrease in its value resulted in a lesser influence on the system's cellular performance. This study provides a novel cross-scaled approach to analyzing sensitivities of computational model parameters and proposes its application to interdisciplinary biomarker studies.

# 1. INTRODUCTION

In the United States, more than 160,000 people die every year of lung cancer, more than breast, colon and prostate cancers combined, and non-small cell lung cancer (NSCLC) accounts for 80% of them (Jemal et al., 2007). Epidermal growth factor receptor (EGFR) is mutated and overexpressed in NSCLC (Hirsch et al., 2003; Paez et al., 2004). A number of different EGFR-related computational models have been developed with an emphasis on explaining signal-response relationships between the binding of epidermal growth factor (EGF) to EGFR and the activation of downstream molecules (Hatakeyama et al., 2003; Kholodenko et al., 1999; Schoeberl et al., 2002). While these models have made successful predictions about the role of different molecular processes in the EGFR signaling cascade, they are limited to providing a qualitative examination of the underlying network properties and the cellular responses they trigger. Therefore, using these models alone, it is difficult to generate a direct quantitative and mechanistic understanding of diverse cellular functions such as cell invasion, proliferation, migration and adhesion within NSCLC. Furthermore, developing predictive models of human disease requires knowledge of different biological levels, including activities within molecular pathways, cells, tissues, organs, and even the entire organism, integrated together to help prioritize therapeutic targets and design clinical trials (Butcher et al., 2004).

The process of model building and experimental validation is expected to be iteratively performed (Di Ventura et al., 2006). To provide more useful knowledge in driving new experiments and generating hypotheses for cancer therapy, signaling events critical to determining the output behavior of a model must be identified (Swameye et al., 2003). These studies are also a major focus of current research in systems biology (Kitano, 2002). Sensitivity analysis has been widely accepted as a useful tool to systematically identify specific perturbations that have significant effects on system behavior, especially when it is not possible or practical to conduct numerous experiments on the living system itself (Frey and Patil, 2002; van Riel, 2006). A sensitivity analysis investigates the effects on the output behavior of a biological system by varying a fixed set of governing parameters or by varying possible combinations of parameters within their expected ranges. In general, system parameters in a signaling pathway model include initial component concentrations and reaction rate constants, both of which can be experimentally measured or inferred to construct the model (van Riel, 2006).

To date, a very large number of modeling efforts involving sensitivity analysis have focused mainly on the signaling pathway scale within cells (Bentele et al., 2004; Cho et al., 2003; Lee et al., 2003; Liu et al., 2005; Mahdavi et al., 2007; Martin and Buckland-Wright, 2004; Zhang and Rundell, 2006; Zi et al., 2005). Some practical applications on robustness analysis of a system, biomarker selection, and drug efficacy evaluation have also been provided (de Pillis et al., 2005; El-Samad et al., 2005; von Dassow et al., 2000), which demonstrate the successful extension of the technique. In *cancer systems biology*, some cellular level models have attempted sensitivity analysis of different system parameters, e.g. vasculature of tumors (Wijeratne and Hoo, 2007) and radiation-induced leukemia (Shuryak et al., 2006). However, to understand the quantitative, dynamical properties of a complex biological system such as cancer, analyzing the sensitivity of parameters at only the molecular or cellular level is not sufficient. Because cancer cells react and respond to, and biological processes take place in, heterogeneous and highly structured biochemical environments (Di Ventura et al., 2006), it is necessary to examine the effect of perturbations in parameters which span both molecular and cellular levels, and beyond.

We previously developed a two-dimensional (2D) simulation model integrating both molecular and cellular levels to examine multicellular dynamics in NSCLC (Wang et al.). An EGFR-ERK signaling transduction pathway specific to NSCLC was proposed and the impact of the change in extrinsic chemotactic stimulus on tumor expansion rate was tested. Here, we use this modeling platform and employ sensitivity analysis to identify critical pathway components and reaction steps and to further test the impact of parameter perturbations on the model output behavior. Like most of the previous sensitivity analysis studies established for signaling pathways (Bentele et al., 2004; Ihekwaba et al., 2004; Lee et al., 2003; Liu et al., 2005; Zhang and Rundell, 2006; Zi et al., 2005), a sensitivity coefficient is applied to our analysis as an index to evaluate the sensitivity of a parameter to the model. However, unique to this study, we utilize the tumor expansion rate—a cellular or microscopic level behavior—as the main, *cross-scale* biological response of the model. While confirming the overall robustness of the model, we successfully identified three critical pathway components and eleven critical reaction steps, and suggest several potential biomarkers that warrant further experimental follow up.

## 2. METHODS

### 2.1 NSCLC Simulation Model

Our previously developed 2D NSCLC model is again employed as the simulation platform in this study; therefore, we will only briefly introduce the concept as well as some key development methods of the model. **Supplementary Figure 1** shows the NSCLC-specific signaling transduction pathway and includes the biochemical reactions that we have previously proposed (Wang et al.). The model consists of ordinary differential equations composed of 20 components downstream of EGF stimulation and 38 corresponding rate constants. Detailed chemical reactions, including rate constants and initial concentrations of components, are described in **Supplementary Tables 1** and **2**. Two of the components, phospholipase Cγ (PLCγ) and extracellular signal-regulated kinase (ERK), are employed to determine two phenotypic traits (proliferation, migration) by comparing their rates of change (ROC) of concentration with corresponding thresholds. The following cellular phenotypic decision algorithm is then applied to the model: a cell will 1) continue to exhibit its previous phenotype if neither of the ROCs of the two components exceed their corresponding thresholds, 2) migrate if only $ROC_{PLC}$ exceeds the threshold of PLCγ, 3) proliferate if only $ROC_{ERK}$ exceeds the threshold of ERK, and 4) in the case that both of the ROCs exceed their corresponding thresholds, migrate if following *Rule A* (migration advantage rule) or proliferate if following *Rule B* (proliferation advantage rule). It should be noted that Rule A and Rule B are artificial rules that we proposed in the absence of any specific experimental data currently available.

Tumor growth dynamics are investigated in a virtual 2D micro-environment with a discrete lattice containing 200 x 200 grid points. A blood vessel representing 'nutrient' source is located at (150,150) and a number of 7 x 7 NSCLC cells are initially positioned in the centre of the lattice. When the first cell reaches the nutrient source, the simulation run is terminated and the elapsed time steps are used as a measure of tumor expansion rate. Three external chemical diffusive cues, EGF, glucose, and oxygen levels, are incorporated into the model and are continuously updated throughout the simulation process. Each grid point within the lattice is assigned a concentration profile determined by these three external cues by means of normal distribution. The nutrient source maintains the highest value of the three cues implicating that it is the most attractive location for the chemotactically acting tumor cells. One of the important features of the multiscale model is that each cell encompasses a self-maintained molecular interaction network. The simulation system records the molecular composite profile for each cell at every time step in order to determine the cell's phenotype for the next step. This essential algorithm establishes the connection between molecular and cellular levels. **Figure 1** shows the cellular phenotypic decision process between two typical time steps. According to our previous findings, Rule A led to a more spatially aggressive tumor with a faster tumor expansion rate than that caused by Rule B. A typical cell expansion pattern conducted using the reference parameter values is shown in **Fig. 2**. Tumor cells are seen to move toward the nutrient source in the NSCLC model. Having a clinical perspective in mind, our interests focus on the more aggressive tumors; hence, sensitivity analysis is conducted on the NSCLC model utilizing Rule A.

**2.2 Sensitivity Analysis**

The method used in the present study belongs to the *local* sensitivity analysis category which is mainly used to evaluate the contribution of individual parameters to the overall performance of a system (Ihekwaba et al., 2004). Alternatively, the *global* sensitivity analysis category encompasses methods that can define the relative importance of parameters related to a system. These analyses, however, require more computational resources and the use of sampling methods to generate random sets of parameter values for simulations. Local and global sensitivity analysis methods may be individually appropriate for different systems depending on the purpose of their implementation, and a recent comparison study showed rankings of the two types (Zhang and Rundell, 2006). In our case, as an initial analysis of the NSCLC model, local sensitivity analysis is a reasonable method of investigation because 1) we are able to perform more detailed analysis on specific parameters that are of particular interest, and 2) local sensitivity analysis can be applied to both linear and non-linear systems (Ingalls and Sauro, 2003).

Within our NSCLC model, molecular and cellular activities are inextricably linked: biological responses at the cellular level, such as migration and proliferation, are determined by examining the change of pathway component concentrations at the molecular level. This design allows us to perform *cross-scale analysis* on the model. We use a sensitivity coefficient as an index to evaluate the effects of perturbations of individual parameter values on the overall system outcome. The coefficient is defined by the following equation (Rabitz et al., 1983):

$$S_p^M = \frac{\delta M / M}{\delta p / p} \qquad (1)$$

where *p* represents the parameter that is varied in a simulation and *M* represents the response of the system; *δM* is the change in *M* due to *δp*, the change in *p*. In our case, *M* corresponds to the tumor expansion rate, i.e. elapsed simulation time steps, while $p$ corresponds to any of the individual parameters, including pathway components and reaction rate constants. The bigger the absolute value of sensitivity coefficient, $|S_p^M|$, the more sensitive is the given parameter. $|S_p^M| > 1$ implies that changes in parameter $p$ may have a significant effect on tumor expansion rate $M$. It is worth noting that, using Eq. (1) has certain drawbacks as a result of adopting the expansion rate as $M$, because, as an extreme example, a cell cannot reach the nutrient source in one step due to the settings of the model, but $p$ can be set to any positive values in a simulation run. Therefore, $p$ must be varied within a finite range, and the accuracy of sensitivity identification decreases as the absolute value of $\delta p / p$ increases. That is, the significance of sensitivity coefficients obtained using an unrealistically large change in value of a given parameter should not be overestimated in determining whether the parameter is critical to the model.

In our analysis, two conditions result in a positive $S_p^M$: an increased expansion rate with decreasing levels of a parameter, and a decreased expansion rate with increasing levels of a parameter. For example, if the expansion rate is increased (which results in decreased elapsed time steps, e.g., from $M_1$ to $M_2$, where $M_2 < M_1$) and a certain parameter value is decreased from $p_1$ to $p_2$, where $p_2 < p_1$, then $\delta M$ is negative and $\delta p$ is negative, and by using equation 1, the resultant $S_p^M$ will be positive. Similarly, an increased expansion rate with increasing levels of a parameter, and a decreased expansion rate with decreasing levels of a parameter result in a negative $S_p^M$. That is, the sensitivity coefficient is an evaluation index for the parameter's sensitivity to the model and does not directly indicate the tumor's expansion rate.

The reference parameter values are taken from the literature and are summarized in **Tables 1** and **2** which list the key components selected from the pathway and the 38 rate constants, respectively. The initial concentrations of the other components (see **Supplementary Fig. 1**) are set to zero. To explore critical parameters, the sensitivity coefficient is calculated using Eq. (1) for each parameter set, perturbing only one parameter and keeping others fixed at their reference values. Variations of each individual parameter should be limited to an expected range; in an effort to cover the entire range of possible, albeit not probable, fluctuations *in vivo,* we investigated an extensive space for each parameter (while remaining mindful of the aforementioned limitations of the technique at very large variations).

## 3. RESULTS

The variation ranges for individual parameters, shown in **Tables 1** and **2**, were set from 0.9-fold (a 10% decrease) to 100-fold of corresponding reference values (however, variation ranges are usually less than 100-fold (Calabrese, 2005)). Each of the variations in the parameters was used as the only change of input when running a simulation, and all other parameters were held fixed at their reference values. This process was repeated for all parameter values, and the resulting sensitivity coefficients were compiled for further analysis.

### 3.1 Critical Components

**Figure 3** illustrates the sensitivity coefficients of each pathway component with respect to the variations listed in **Table 1**. Protein kinase C (PKC), mitogen-activated protein kinase kinase (MEK) and ERK are most sensitive to the model, as seen by their sensitivity coefficients near one, followed by less-sensitive EGFR, PLCγ and Raf. Therefore, PKC, MEK and ERK are *critical* components of the pathway (currently) implemented. The peak maxima of sensitivity coefficient plots of the three critical components occur at a variation of 1.1-fold of their corresponding reference values. At a variation of 2.0-fold, their sensitivity coefficients decrease dramatically to a small value, meaning the expansion rate did not deviate much from that of the reference value, indicating that the expansion of the system is no longer subject to the increase in these parameter values. We call such ranges of variation that express the sensitive property of a model component the "critical area". In our study, the critical areas for the three components are variations between 1.1- and 2.0-fold of reference values.

To gain more insight into how small changes in the concentration values of the three components will affect the system expansion rate, we performed more detailed analyses on PKC, MEK, and ERK. We varied concentration values from 1.0- (but not including 1.0) to 1.2-fold by an incremental increase of 0.01, and from 1.2- to 2.0-fold by an incremental increase of 0.1 within each component's critical area. The resulting sensitivity coefficients are shown in **Fig. 4**. It can be seen that sensitivity coefficients of all the three components drop gradually as the variation increases.

### 3.2 Critical Steps

We conducted further investigations to identify critical steps within the model, calculating sensitivity coefficients for each of the 38 rate constants. **Table 2** lists for each rate constant the sensitivity coefficient (whose absolute values is the largest obtained for the rate constant) and the corresponding variation. We find twelve sensitive rate constants (whose maximum absolute sensitivity coefficients are equal to 0.9), corresponding to reaction steps: $v_5$, $v_6$, $v_7$, $v_9$, $v_{10}$, $v_{11}$, $v_{12}$, $v_{14}$, and $v_{15}$ (see **Table 2**), which are considered to be more critical than others. Again, more detailed analyses of these steps were carried out by examining smaller changes in variation. Because maximum absolute values of sensitivity coefficients of some rate constants occurred at 0.9-fold variation, we determined the critical area from 0.1- to 1.0-fold for these rate constants, and from 1.0- to 2.0-fold for the remaining constants. The resulting sensitivity analyses are shown in **Fig. 5**. From these detailed tests, the critical reaction steps are further divided into two groups: a critical group (CG) in which the maximum absolute value of sensitivity coefficients of a reaction step is less than 5.0 and greater than 1.0, and a highly critical group (HCG) if the maximum absolute value is greater than or equal to 5.0. Accordingly, reaction steps, $v_6$, $v_7$, $v_9$, $v_{10}$ and $v_{11}$ belong to CG, and $v_{12}$, $v_{14}$ and $v_{16}$ belong to HCG. Because the maximum absolute value of sensitivity coefficients of $k_{5-b}$ (corresponding to $v_5$, see **Table 2**) is still 0.9 after further study, we consider it to be non-critical. Consistent with our previous finding on critical components, $v_{10}$ (CG), $v_{12}$ (HCG), and $v_{16}$ (HCG) are reaction steps directly related to PKC, MEK, and ERK, respectively.

## 4. DISCUSSION

When conducting parameter analysis of a computational cancer model as a means to make correct predictions and to guide treatment, focusing on the molecular *or* the cellular level *only* may not be suitable because cells respond to heterogeneous and highly structured biochemical environments (Di Ventura et al., 2006). In a signaling pathway model which functions only on the molecular level, it is likely that a wide range of parameters will fit the ultimate signal events (i.e., will lead to the same final solution), but they may lead to different cellular responses when examining the pathway in different cellular environments. Taking the EGFR-ERK signaling pathway as an example, **Fig. 6** schematically illustrates the possible difference between molecular and cellular scale responses dependent on microenvironmental conditions ('context-specific' outcome). On the other hand, only environmental (high-level) parameters can be analyzed in cellular models, which do not always reveal the mechanisms underpinning the observed phenomena. Our previously developed *multiscale* model for NSCLC investigated the effects of intracellular events on cellular responses (Wang et al.). Therefore, using this previously developed *in silico* platform, we were able to conduct this study here to assess the influence of molecular level parameters on a cellular level response. The model response utilized was not a behavior of output signals or signal activation patterns, but rather the tumor expansion rate (a phenotypic behavior at the cellular level), which was influenced by both signaling components and tumor growth environment. To our knowledge, this is the first time to investigate parameter sensitivities of a cancer model, *across* different biological levels.

The analysis results confirmed the *robustness* of the previously developed NSCLC model. The model was capable of tolerating up to 100-fold variation in most parameters, facilitating adaptation of the model to different cancer systems without much system-specific readjustment of parameters. However, some components such as PKC, MEK and ERK (see **Fig. 3**) were found to be *critical* to the stability of the model in terms of their influence on the selected microscopic performance evaluator. This means that minor quantitative variations in any such sensitive parameter, within its critical area, led the model to respond drastically with regards to the tumor's spatio-temporal expansion rate. However, the fact that the critical area for these sensitive parameters was rather small further supports the notion of a tightly regulated signaling network. While we are aware of the fact that the pathway implemented is incomplete and its context over-simplified, we nonetheless argue that for the current setup these critical components can be understood as *biomarkers* because, as their values are altered, the model produces distinct cellular responses which may lead to different characteristic disease phenotypes (Frey and Patil, 2002). The importance of MEK and ERK determined in our works here is in agreement with another control analysis study on a more complex kinetic model of EGF-induced MAPK signaling (Hornberg et al., 2005). We also note that many experimental and pharmaceutical studies have demonstrated the substantial potential of PKC as a biomarker, especially in human breast cancer, colon cancer, and NSCLC (Bae et al., 2007; Davidson et al., 1998; Green et al., 2006; Nagashima et al., 2007), as well as the use of ERK in NSCLC (Han et al., 2005; Vicent et al., 2004). While MEK has not been a large focus in biomarker studies, it does play a significant role in the generic MAPK cascade (Wakeling, 2005). The activation and inhibition of MEK result in different signal output in terms of strength and duration (Allen et al., 2003; Gollob et al., 2006), therefore, possible use of MEK as a biomarker (as a profile, in conjunction with PKC and ERK, or separately) should be experimentally investigated. Intriguingly, recent results in other solid cancers, i.e. gliomas, implicating MEK in the phenotypic decision process, seem to corroborate our findings (Demuth et al., 2007).

Some pathway components failed to demonstrate critical impact in the current setup, although their significance for a variety of cell responses in cancer has been experimentally confirmed (such as PLCγ for tumor cell motility (Mouneimne et al., 2004; Piccolo et al., 2002), Raf for tumorigenesis (Gollob et al., 2006) and cell differentiation (Hirsch et al., 2006)) which originally led to their inclusion in our *in silico* model. Because PLCγ was used as the key regulating component for exploring a migratory phenotypic switch, one would have expected it to have greater control over the behavior of the model. A closer look reveals this is a result largely of the selection of the designated network performance evaluator (the system outcome). After all, choosing an appropriate output target is a challenge in calculating context-specific sensitivity of parameters (Aldridge et al., 2006). In this study, we chose the tumor's expansion rate to be the system outcome $M$ in Eq. (1) to suit our clinical perspective. Furthermore, the model examined the *rate of change* of PLCγ, not the PLCγ concentration, in order to process a migration decision, and thereby even if the concentration of PLCγ is higher, its rate of change may not be. However, this result also confirmed that the assignment of molecular threshold steps does *not* necessarily predetermine their weight on the phenotypic behavior they induce across scales, and that different thresholds (PLCγ versus ERK) can have distinctively different impact.

The most interesting finding in our analysis results is that, for both initial component concentrations and reaction rate constants, sensitivity coefficient plots of the sensitive parameters revealed similar patterns. Each of the sensitive parameters has a critical area, and if a variation occurred in that area, the varying parameter values had strong influence on the cancer system's expansion rate which served as our microscopic "read-out". All sensitivity coefficients decreased with an increase of variation within the critical area. Interestingly, here a very small variation in a parameter reference value resulted in a large change on the final output of the model, but conversely a relatively large change did not substantially alter model output. Together, this suggests a tightly coupled and highly efficient sub-cellular information processing system where even minimal modulations in signal strength are sufficient to elicit major phenotypic changes. Experimentally, this phenomena has been commonly reported with regards to dose-response relationship of human tumor cell lines (see (Calabrese, 2005) for a review). For example, in testing responses to transforming growth factor-β (TGF-β) stimulation, a low TGF-β concentration results in increased cellular proliferation in prostate carcinoma cell lines (Ritchie et al., 1997). In NSCLC cell lines, insulin-like growth factor-I (IGF-I) significantly inhibits cell proliferation at higher concentrations (Kodama et al., 2002) Additionally, some *in silico* pathway models also predict the aforementioned sensitivity response in final molecular signaling events (Kharait et al., 2007; Liu et al., 2005; Mahdavi et al., 2007).

## 5. CONCLUSION & FUTURE WORKS

In summary, using a multiscale NSCLC model, this paper presents a innovative approach to investigate parameter sensitivities of a cancer model by taking *both* molecular *and* cellular levels into account. While, overall, the model displayed robustness to relatively large fluctuations, some parameters had more impact on the system's multicellular performance than others. A small variation in the reference value of any critical parameter, including three pathway components and eleven reaction steps, resulted in a relatively large change in the multicellular output of the model. In the future, we plan on simulating a three-dimensional biochemical environment at the cellular level which should help provide a more accurate representation of the *in vivo* situation. At the molecular level, while a computational model cannot be a biological representation in every detail, integration of other pathways of relevance into the current NSCLC model, such as e.g.

PI3K/PTEN/AKT (Vivanco and Sawyers, 2002) and TGF-β (Akhurst and Derynck, 2001), will be utilized. Global sensitivity analysis will then be exploited in investigating the possibilities of combined effects of parameter variations for a wide range of possible values for all parameters simultaneously. While comprehensive analysis strains computing resources, random sets of parameter values for simulations can be generated more easily with advanced sampling methods and will benefit model analysis. Although still at the beginning, we nonetheless believe that this *cross-scale sensitivity* study provides a novel, useful method in exploring and ranking biomarkers, and, more generally, supports the use of such multiscale models in interdisciplinary cancer research.

## ACKNOWLEDGEMENTS

This work has been supported in part by NIH grant CA 113004 and by the Harvard-MIT (HST) Athinoula A. Martinos Center for Biomedical Imaging and the Department of Radiology at Massachusetts General Hospital.

# CAPTIONS

**Figure 1:** Phenotypic decision process for a cancer cell between two typical time steps.

**Figure 2:** A typical cell expansion pattern in the NSCLC model. Proliferative cells are labeled in *dark blue*, migratory cells in *red*, quiescent cells in *green* and dead cells in *grey*.

**Figure 3:** The sensitivity coefficients of each selected pathway component with respect to the variations listed in **Table 1**.

**Figure 4:** The sensitivity coefficients of most sensitive pathway components (PKC, MEK and ERK) with respect to the variations from 1.0- to 1.2-fold by an incremental increase of 0.01 (left panel) and from 1.2- to 2.0-fold by an incremental increase of 0.1 (right panel).

**Figure 5:** Plots of sensitivity coefficients for sensitive rate constants. In **(a)**, seven rate constants are shown for which the critical area is between 1.0- and 2.0-fold; their values varied from 1.0- to 1.2-fold by an incremental increase of 0.01 (first and third columns) and from 1.2- to 2.0-fold by an incremental increase of 0.1 (second and fourth columns). In **(b)**, five rate constants are shown for which the critical area is between 0.1 and 1.0; their values varied from 0.1- to 0.8-fold by an incremental increase of 0.1 (first and third columns) and from 0.8- to 1.0-fold by an incremental increase of 0.01 (second and fourth columns).

**Figure 6:** Schematic illustration of how perturbations (a, b, c, ..., n) in one or more sub-cellular parameters, *p*, can yield the same dynamic response at the molecular level, yet may lead to distinct responses at the cellular level dependent on the microenvironment.

**FIGURES & TABLES**

**FIGURE 1.**

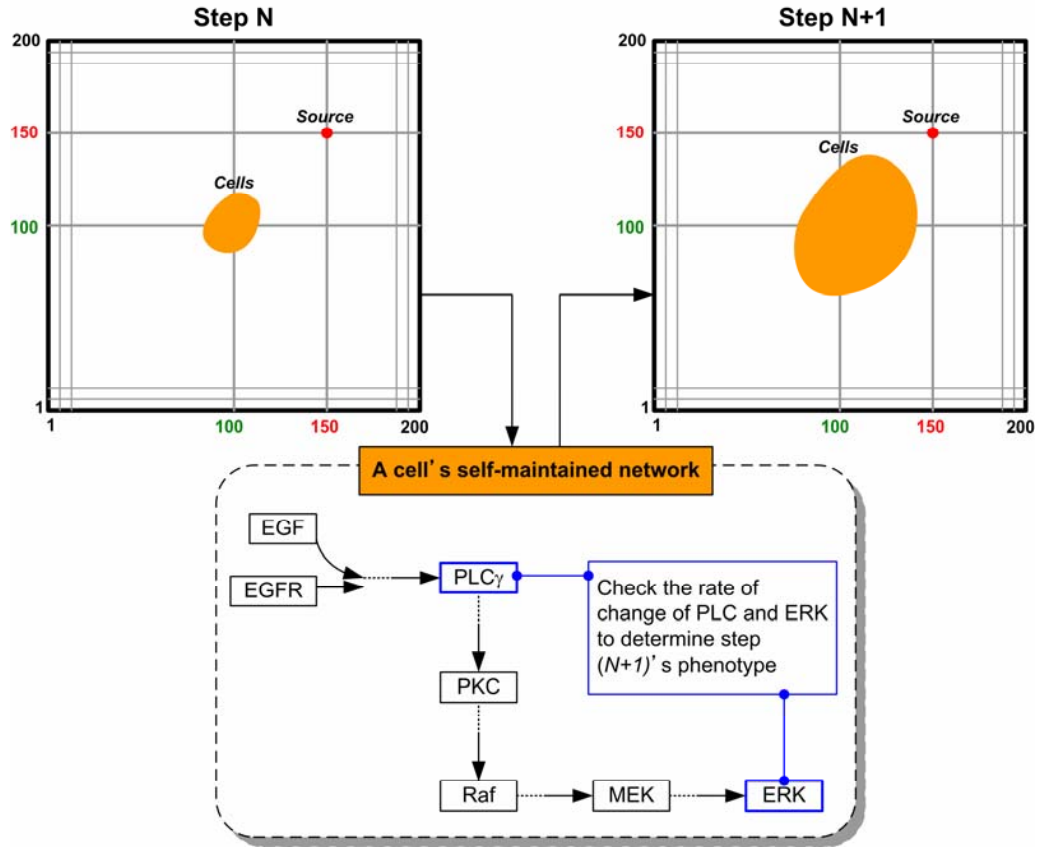

**FIGURE 2.**

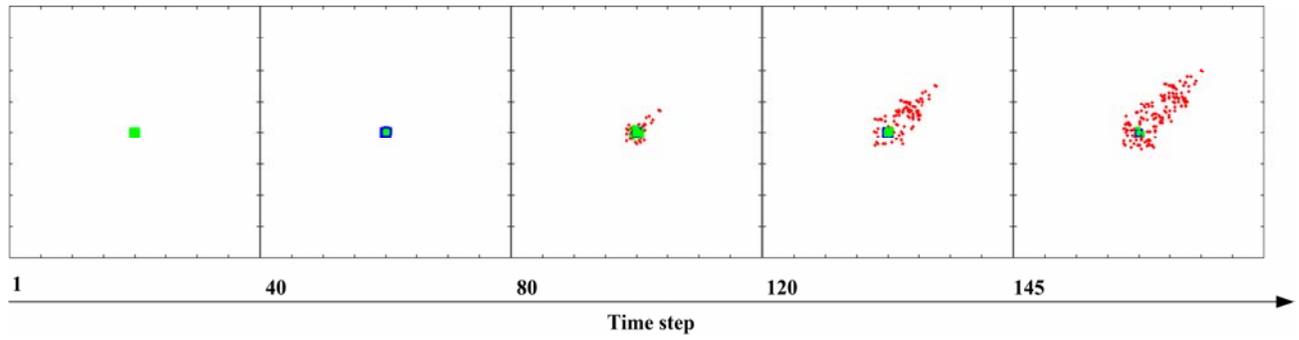

**FIGURE 3.**

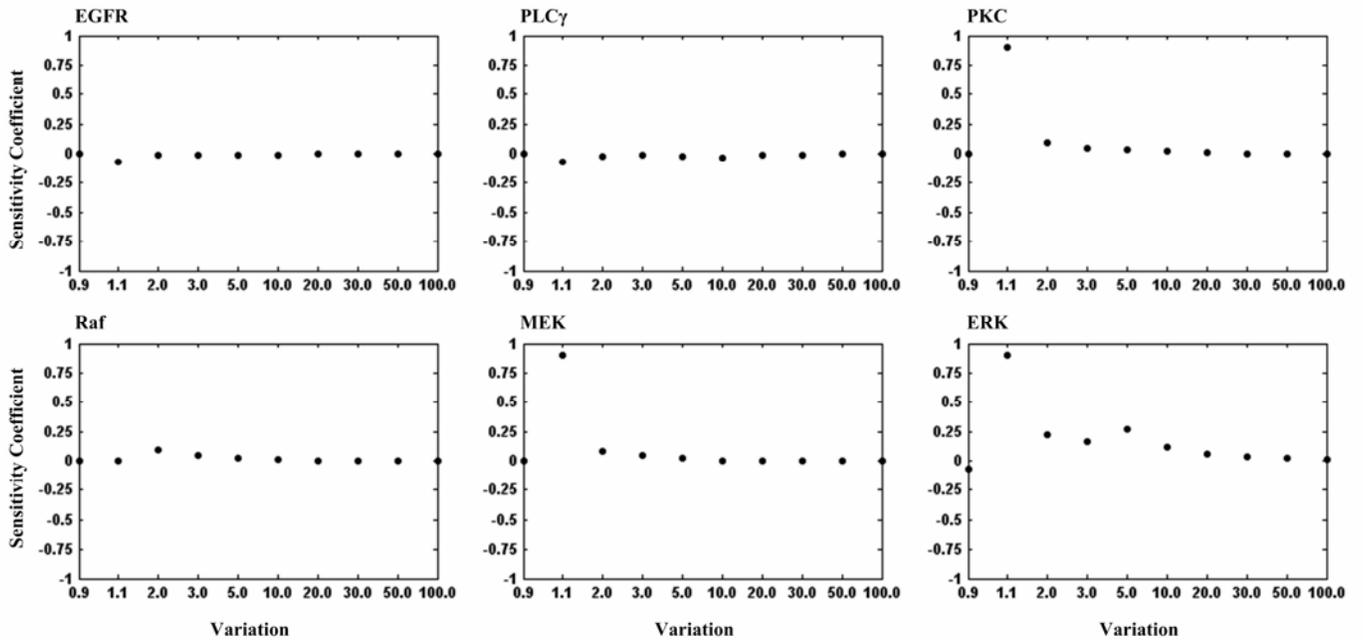

**FIGURE 4.**

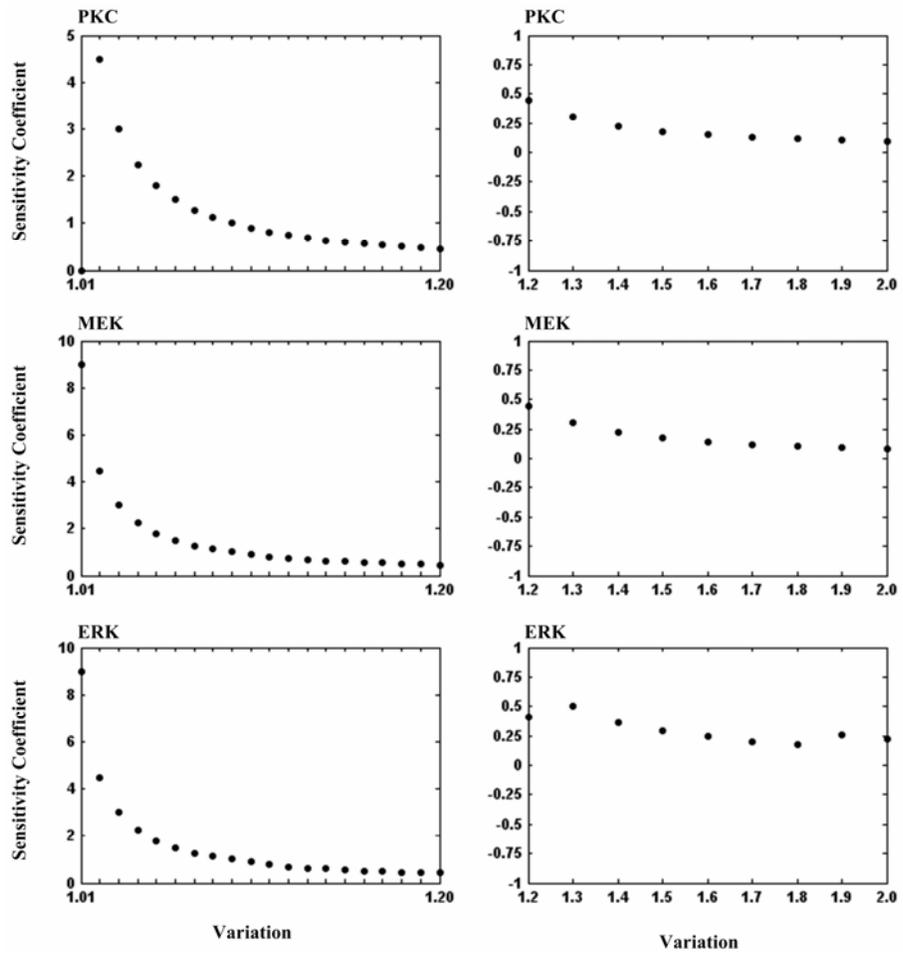

Z. Wang et al.: *Sensitivity Analysis of a NSCLC Model***FIGURE 5.**

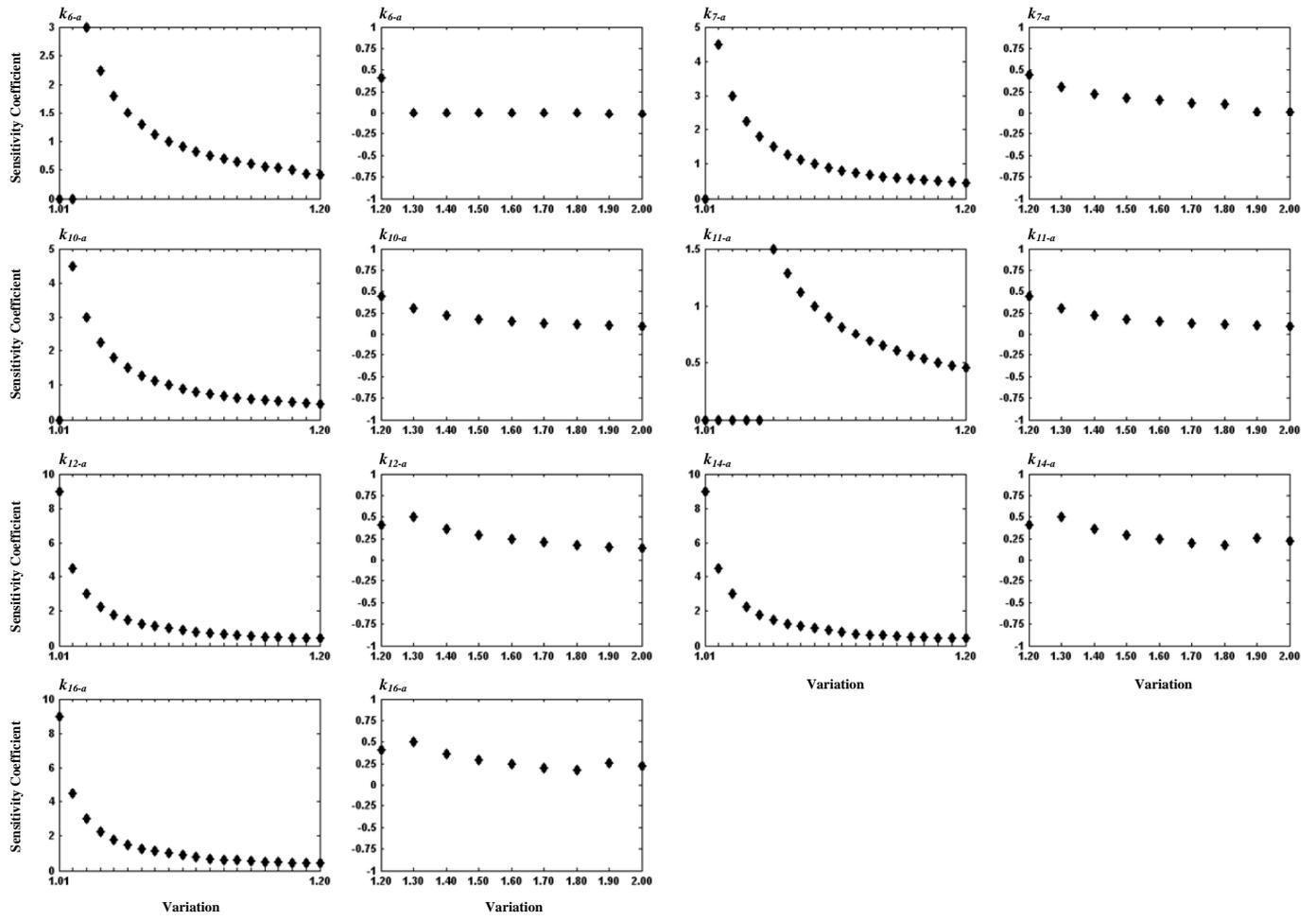

(a)

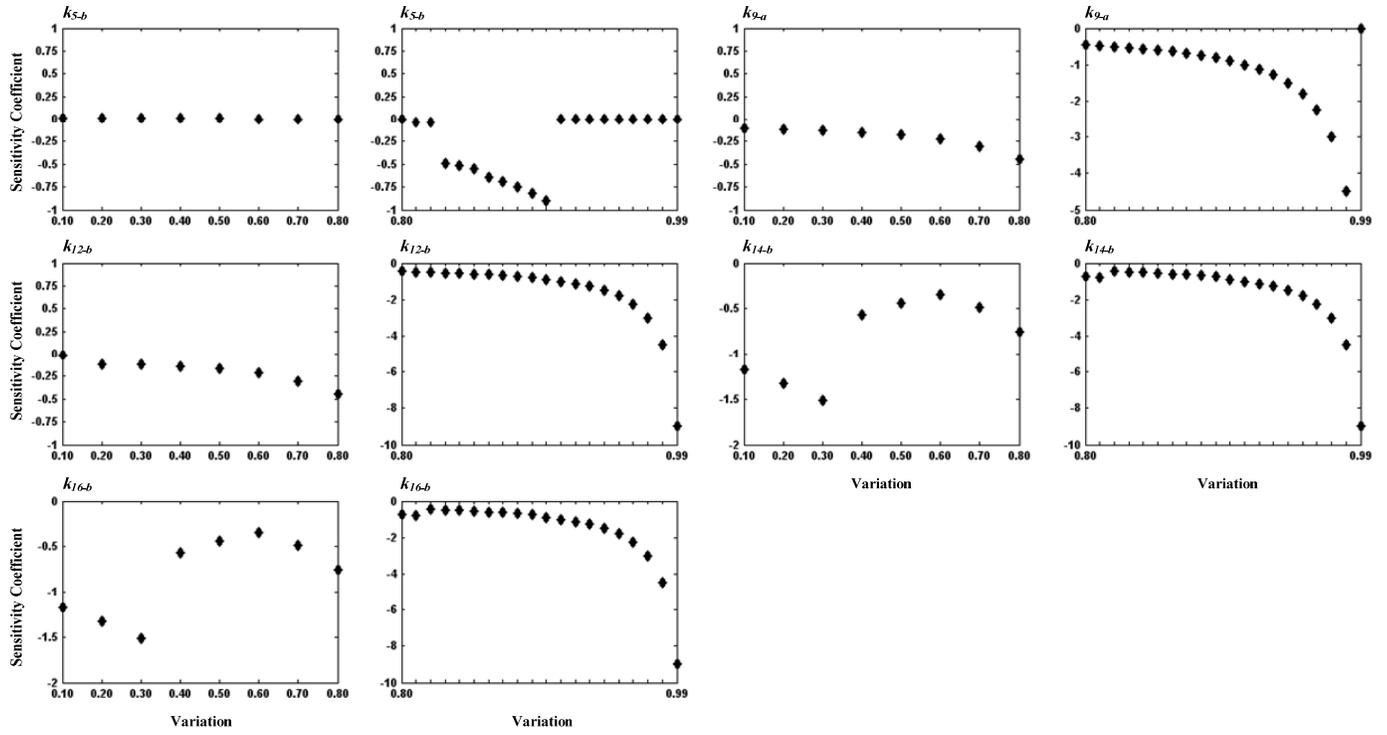

**(b)**

**FIGURE 6.**

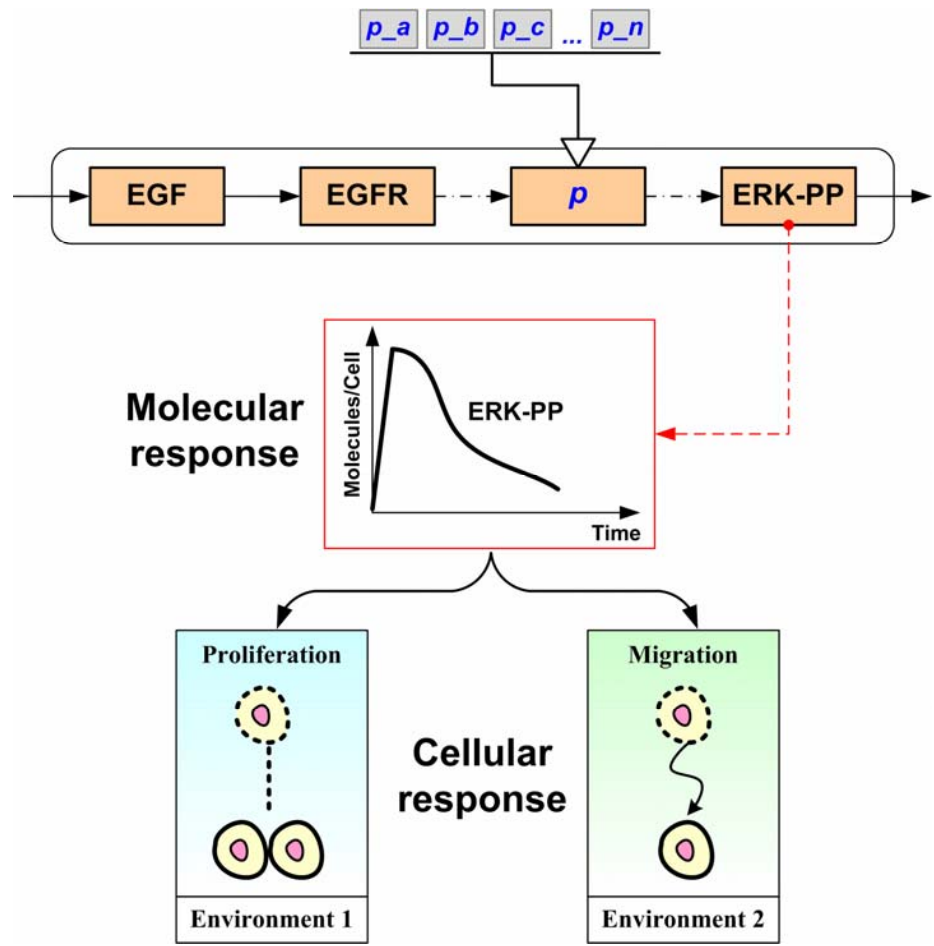

**TABLE 1.** Reference values and variations of individual pathway components. Values are taken from the literature (Hatakeyama et al., 2003; Kholodenko et al., 1999; Schoeberl et al., 2002).

| Pathway Component | Reference initial concentration [nM] | Range of variation |
|---|---|---|
| EGFR | 80 | |
| PLCγ | 10 | |
| PKC | 10 | 0.9, 1.0, 2.0, 3.0, 5.0, 10.0, 20.0, 30.0, 50.0, 100.0 |
| Raf | 100 | |
| MEK | 120 | |
| ERK | 100 | |

**TABLE 2.** Reference values and variations of association and dissociation rate constants. Values are taken from the literature (Hatakeyama et al., 2003; Kholodenko et al., 1999; Schoeberl et al., 2002).

| Reaction number | Rate constant | Reference value | Range of variation | Maximum absolute value of sensitivity coefficient | Variation |
|---|---|---|---|---|---|
| $v_1$ | $k_{1-a}$ | 0.003 | | -0.07 | 1.1 |
| | $k_{1-b}$ | 0.06 | | -0.83 | 0.9 |
| $v_2$ | $k_{2-a}$ | 0.01 | | -0.07 | 1.1 |
| | $k_{2-b}$ | 0.1 | | -0.83 | 0.9 |
| $v_3$ | $k_{3-a}$ | 1 | | -0.07 | 1.1 |
| | $k_{3-b}$ | 0.01 | | 0 | – |
| $v_4$ | $k_{4-a}$ | 450 | | 0.44 | 3 |
| | $k_{4-b}$ | 50 | | -0.07 | 1.1 |
| $v_5$ | $k_{5-a}$ | 0.06 | | -0.07 | 1.1 |
| | $k_{5-b}$ | 0.2 | | -0.9 | 0.9 |
| $v_6$ | $k_{6-a}$ | 1 | | 0.9 | 1.1 |
| | $k_{6-b}$ | 0.05 | | 0 | – |
| $v_7$ | $k_{7-a}$ | 0.3 | | 0.9 | 1.1 |
| | $k_{7-b}$ | 0.006 | | 0 | – |
| $v_8$ | $k_{8-a}$ | 1 | | 0 | – |
| | $k_{8-b}$ | 100 | | 0 | – |
| $v_9$ | $k_{9-a}$ | 1 | | -0.9 | 0.9 |
| | $k_{9-b}$ | 0.03 | | 0.09 | 2 |
| $v_{10}$ | $k_{10-a}$ | 0.214 | 0.9, 1.0, 2.0, 3.0, 5.0, 10.0, 20.0, 30.0, 50.0, 100.0 | 0.9 | 1.1 |
| | $k_{10-b}$ | 5.25 | | 0 | – |
| $v_{11}$ | $k_{11-a}$ | 4 | | 0.9 | 1.1 |
| | $k_{11-b}$ | 64 | | 0 | – |
| $v_{12}$ | $k_{12-a}$ | 3.5 | | 0.9 | 1.1 |
| | $k_{12-b}$ | 317 | | -0.9 | 0.9 |
| $v_{13}$ | $k_{13-a}$ | 0.058 | | 0 | – |
| | $k_{13-b}$ | 2200 | | 0 | – |
| $v_{14}$ | $k_{14-a}$ | 2.9 | | 0.9 | 1.1 |
| | $k_{14-b}$ | 317 | | -0.9 | 0.9 |
| $v_{15}$ | $k_{15-a}$ | 0.05 | | 0 | – |
| | $k_{15-b}$ | 60 | | 0.09 | 2 |
| $v_{16}$ | $k_{16-a}$ | 9.5 | | 0.9 | 1.1 |
| | $k_{16-b}$ | $1.46 \times 10^5$ | | -0.9 | 0.9 |
| $v_{17}$ | $k_{17-a}$ | 0.3 | | 0 | – |
| | $k_{17-b}$ | 160 | | 0.09 | 2 |
| $v_{18}$ | $k_{18-a}$ | 16 | | 0 | – |
| | $k_{18-b}$ | $1.46 \times 10^5$ | | 0 | – |
| $v_{19}$ | $k_{19-a}$ | 0.27 | | 0 | – |
| | $k_{19-b}$ | 60 | | 0 | – |

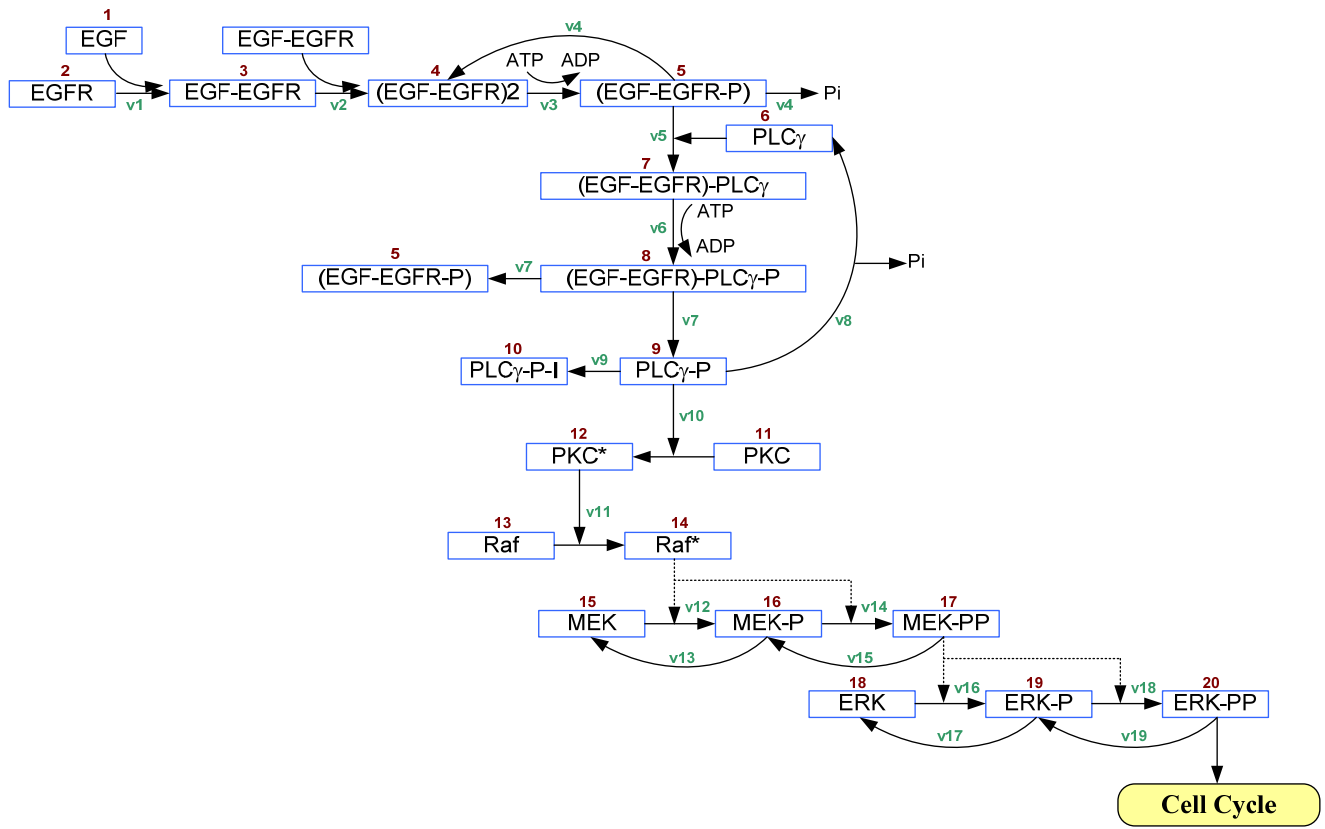

**Supplementary Figure 1.** Kinetic model of the NSCLC-specific EGFR signaling pathway, modified from (Wang et al.). The arrows represent the reactions specified in **Supplementary Tables 1** and **2** and characterized by reaction rates v1–v19 (*green* numbers). Each pathway component is identified by a specific number (*brown* numbers).

**Supplementary Table 1.** Kinetic equations and initial concentrations.

| Reactant | Molecular variable | Initial concentration [nM] | Ordinary Differential Equation |
|---|---|---|---|
| $X_1$ | EGF | to be varied | $d(X_1)/dt = -v_1$ |
| $X_2$ | EGFR | 80 | $d(X_2)/dt = -v_1$ |
| $X_3$ | EGF-EGFR | 0 | $d(X_3)/dt = v_1 - 2v_2$ |
| $X_4$ | (EGF-EGFR)2 | 0 | $d(X_4)/dt = v_2 + v_4 - v_3$ |
| $X_5$ | EGF-EGFR-P | 0 | $d(X_5)/dt = v_3 + v_7 - v_4 - v_5$ |
| $X_6$ | PLCγ | 10 | $d(X_6)/dt = v_8 - v_5$ |
| $X_7$ | EGF-EGFR-PLCγ | 0 | $d(X_7)/dt = v_5 - v_6$ |
| $X_8$ | EGF-EGFR-PLCγ-P | 0 | $d(X_8)/dt = v_6 - v_7$ |
| $X_9$ | PLCγ-P | 0 | $d(X_9)/dt = v_7 - v_8 - v_9 - v_{10}$ |
| $X_{10}$ | PLCγ-P-I | 0 | $d(X_{10})/dt = v_9$ |
| $X_{11}$ | PKC | 10 | $d(X_{11})/dt = -v_{10}$ |
| $X_{12}$ | PKC* | 0 | $d(X_{12})/dt = v_{10} - v_{11}$ |
| $X_{13}$ | Raf | 100 | $d(X_{13})/dt = -v_{11}$ |
| $X_{14}$ | Raf* | 0 | $d(X_{14})/dt = v_{11} - v_{12} - v_{14}$ |
| $X_{15}$ | MEK | 120 | $d(X_{15})/dt = v_{13} - v_{12}$ |
| $X_{16}$ | MEK-P | 0 | $d(X_{16})/dt = v_{12} + v_{15} - v_{13} - v_{14}$ |
| $X_{17}$ | MEK-PP | 0 | $d(X_{17})/dt = v_{14} - v_{15} - v_{16} - v_{18}$ |
| $X_{18}$ | ERK | 100 | $d(X_{18})/dt = v_{17} - v_{16}$ |
| $X_{19}$ | ERK-P | 0 | $d(X_{19})/dt = v_{16} + v_{19} - v_{17} - v_{18}$ |
| $X_{20}$ | ERK-PP | 0 | $d(X_{20})/dt = v_{18} - v_{19}$ |

**Supplementary Table 2.** Kinetic parameters. Concentrations and the Michaelis-Menten constants ($K_4$, $K_8$, and $K_{11}$–$K_{19}$) are given in [nM]. First- and second-order rate constants are given in [s$^{-1}$] and [nM$^{-1}$ · s$^{-1}$], respectively. $V_4$, $V_8$, and $V_{11}$–$V_{19}$ are expressed in [nM · s$^{-1}$].

| Reaction number | Equation | Kinetic parameter | | Reference |
|---|---|---|---|---|
| $v_1$ | $k_1 \cdot X_1 \cdot X_2 - k_{-1} \cdot X_3$ | $k_1$=0.003 | $k_{-1}$=0.06 | [1] |
| $v_2$ | $k_2 \cdot X_3 \cdot X_3 - k_{-2} \cdot X_4$ | $k_2$=0.01 | $k_{-2}$=0.1 | [1] |
| $v_3$ | $k_3 \cdot X_4 - k_{-3} \cdot X_5$ | $k_3$=1 | $k_{-3}$=0.01 | [1] |
| $v_4$ | $V_4 \cdot X_5 / (K_4 + X_5)$ | $V_4$=450 | $K_4$=50 | [1] |
| $v_5$ | $k_5 \cdot X_5 \cdot X_6 - k_{-5} \cdot X_7$ | $k_5$=0.06 | $k_{-5}$=0.2 | [1] |
| $v_6$ | $k_6 \cdot X_7 - k_{-6} \cdot X_8$ | $k_6$=1 | $k_{-6}$=0.05 | [1] |
| $v_7$ | $k_7 \cdot X_8 - k_{-7} \cdot X_5 \cdot X_9$ | $k_7$=0.3 | $k_{-7}$=0.006 | [1] |
| $v_8$ | $V_8 \cdot X_9 / (K_8 + X_9)$ | $V_8$=1 | $K_8$=100 | [1] |
| $v_9$ | $k_9 \cdot X_9 - k_{-9} \cdot X_{10}$ | $k_9$=1 | $k_{-9}$=0.03 | [1] |
| $v_{10}$ | $k_{10} \cdot X_9 \cdot X_{11} - k_{-10} \cdot X_{12}$ | $k_{10}$=0.214 | $k_{-10}$= 5.25 | Estimate |
| $v_{11}$ | $V_{11} \cdot X_{12} \cdot X_{13} / (K_{11} + X_{13})$ | $V_{11}$=4 | $K_{11}$=64 | [2] |
| $v_{12}$ | $V_{12} \cdot X_{14} \cdot X_{15} / [K_{12} \cdot (1 + X_{16} / K_{14}) + X_{15}]$ | $V_{12}$=3.5 | $K_{12}$=317 | [3] |
| $v_{13}$ | $V_{13} \cdot X_{16} / [K_{13} \cdot (1 + X_{17} / K_{15}) + X_{16}]$ | $V_{13}$=0.058 | $K_{13}$=2200 | [4] |
| $v_{14}$ | $V_{14} \cdot X_{14} \cdot X_{16} / [K_{14} \cdot (1 + X_{15} / K_{12}) + X_{16}]$ | $V_{14}$=2.9 | $K_{14}$=317 | [4] |
| $v_{15}$ | $V_{15} \cdot X_{17} / [K_{15} \cdot (1 + X_{16} / K_{13}) + X_{17}]$ | $V_{15}$=0.058 | $K_{15}$=60 | [4] |
| $v_{16}$ | $V_{16} \cdot X_{17} \cdot X_{18} / [K_{16} \cdot (1 + X_{19} / K_{18}) + X_{18}]$ | $V_{16}$=9.5 | $K_{16}$=1.46 × 10$^5$ | [4] |
| $v_{17}$ | $V_{17} \cdot X_{19} / [K_{17} \cdot (1 + X_{20} / K_{19}) + X_{19}]$ | $V_{17}$=0.3 | $K_{17}$=160 | [4] |
| $v_{18}$ | $V_{18} \cdot X_{17} \cdot X_{19} / [K_{18} \cdot (1 + X_{18} / K_{16}) + X_{19}]$ | $V_{18}$=16 | $K_{18}$=1.46 × 10$^5$ | [4] |
| $v_{19}$ | $V_{19} \cdot X_{20} / [K_{19} \cdot (1 + X_{19} / K_{17}) + X_{20}]$ | $V_{19}$=0.27 | $K_{19}$=60 | [4] |